\documentclass[pra,twocolumn,floatfix,superscriptaddress]{revtex4-1}
\usepackage{dcolumn,amsmath}
\usepackage{graphicx}
\usepackage{bm}
\usepackage{amssymb}
\usepackage{multirow}

\usepackage{amsfonts}
\usepackage{amsmath}
\usepackage{cancel}
\usepackage{xcolor}
\begin{document}
\title{Electric field dependent g factors of YbOH molecule}
\author {Alexander Petrov}\email{petrov\_an@pnpi.nrcki.ru}
\affiliation{Petersburg Nuclear Physics Institute named by B.P. Konstantinov of National Research Centre
"Kurchatov Institute", Gatchina, 1, mkr. Orlova roshcha, 188300, Russia}
\affiliation{St. Petersburg State University, St. Petersburg, 7/9 Universitetskaya nab., 199034, Russia} 

\date{Received: date / Revised version: date}
%
\begin{abstract}{
 YbOH molecule is one of the most sensitive systems for the electron electric dipole moment ($e$EDM) searches. Zeeman splittings of the $e$EDM sensitive levels have significant implications 
 to control and suppress
important systematic effects due to stray magnetic field in experiments for $e$EDM searches.
 The electric-field-dependent g factors of the 
 lowest rotational level of the first excited bending vibrational mode of $^{174}$YbOH are calculated. l-doublet levels with small g factors difference are found and main contributions to the difference are determined. 
} 
\end{abstract}
\maketitle
\section{Introduction}
Measuring the electron electric dipole moment ($e$EDM) is now considered as a most promising test for existence of physics beyond the Standard model \cite{Fukuyama2012,PospelovRitz2014,YamaguchiYamanaka2020,YamaguchiYamanaka2021}. 
With a first excited $v=1$ ($v$ is a vibrational quantum number) bending vibrational mode sensitive to $\mathcal{T},\mathcal{P}$parity-violating ($\mathcal{T}$ -- time reversal, $\mathcal{P}$ -- space parity)
effects, the $^{174}$YbOH has been considering as
a promising candidate for the $e$EDM searches. It has been studied and
discussed in many papers, see e.g.  \cite{Isaev:16, Kozyryev:17, Steimle:2019, maison2019theoretical, Augenbraun_2020, Pilgram:2021, PhysRevA.103.022813, Petrov:2022, Kurchavov2022, Persinger:2023, Jadbabaie_2023, Petrov:2024}  and references therein. 

In the  $e$EDM experiment the energy splitting between levels with opposite direction of the full momentum projections is measured
at the presence of the electric and magnetic fields being parallel or antiparallel to each other. Besides  $e$EDM there is the Zeeman induced contribution to the splitting . Thus insufficient control of the magnetic field is a source of systematic errors in the experiment.
Diatomic molecules with $\Omega$-doublets are very robust against these systematics, since the $\Omega$-doublet structure is arranged in such a way that $e$EDM contribution to the splitting is opposite in two $\Omega-$doublet states, whereas the Zeeman ones have the same sign. Thus the energy splittings in the two $\Omega$-doublet states can be subtracted from each other, which suppresses systematic effects related to the stray magnetic field (and many others systematic effects) but doubles the $e$EDM signal.  The advantages of $\Omega$-doublets for suppression of systematic effects were first proposed in Ref. \cite{DeMille2001}.
As a consequence a great progress in $e$EDM search on HfF$^+$ \cite{newlimit1} and ThO \cite{ACME:18} is closely related to $\Omega-$doubling structure of these molecules.

  The effect of $l$-doubling (see below) in triatomic molecules with linear equilibrium configuration in the first excited bending vibrational mode is quite similar to $\Omega$-doubling one in diatomic molecules. 
  Thus, the existence of  $l$-doublets in the excited $v=1$ bending vibrational mode helps to suppress many systematics and allows to polarize the molecule in rather small electric field \cite{Kozyryev:17, Petrov:2022}. Beyond this in such molecules the sensitivity of the experiments can be strongly enhanced due to laser cooling and increasing the coherence time \cite{Isaev_2017, Kozyryev:17}.
  Recently a quantum control of trapped cooled triatomic molecules for $e$EDM searches was demonstrated \cite{anderegg2023quantum}.
  
  However, the upper and lower states of $l$-doublet (as well as $\Omega$-doublet) have slightly different magnetic g factors, and this difference $\Delta {\rm g}$ depends on the laboratory electric field. Therefore it is clear that understanding the g factor dependence  on electric field is important for understanding possible systematic effects and their control in $e$EDM search experiments.  
   Systematic effects related to magnetic field imperfections are suppressed by a factor of $\sim\Delta {\rm g}/{\rm g}$. 
   
  Previously the g factor dependence on electric field was studied for many diatomics including PbO \cite{Petrov:11}, WC \cite{Lee:13a}, ThO \cite{Petrov:14}, HfF$^+$  \cite{ Petrov:17b, Kurchavov:2020, Kurchavov2021}, RaF\cite{Petrov:2020} and PbF\cite{Baturo:2021}. However, up to now there are no such calculations for linear triatomic molecules. Even the  order of magnitude of $\Delta {\rm g}$ as well as contributions to the difference are unclear currently. In the current work we address the problem 
  and calculated g factors of the $^{174}$YbOH in the ground rotational level of the excited bending vibrational mode.

\section{Method} 
Following Ref. \cite{Petrov:2022, Petrov:2024}  we present our Hamiltonian in molecular reference frame as
\begin{equation}
{\rm \bf\hat{H}} = {\rm \bf\hat{H}}_{\rm mol} + {\rm \bf\hat{H}}_{\rm hfs} + {\rm \bf\hat{H}}_{\rm S} + {\rm \bf\hat{H}}_{\rm Z},
\label{Hamtot}
\end{equation} 
where
\begin{equation}
{\rm \bf\hat{H}}_{\rm mol}=\frac{(\hat{\bf J} -\hat{\bf J}^{e-v} )^2}{2\mu R^2}+\frac{(\hat{\bf J}^{v})^2}{2\mu_{\rm OH}r^2}+ V(\theta)
\label{Hmolf}
\end{equation}
is the molecular Hamiltonian,
$\mu$ is the reduced mass of the Yb-OH system, $\mu_{\rm OH}$ is the reduced mass of the OH, $\hat{\bf J}$ is the total electronic, vibrational, and rotational
angular momentum, $\hat{\bf J}^{e-v} = \hat{\bf J}^{e} + \hat{\bf J}^{v}$ is the electronic-vibrational momentum, $\hat{\bf J}^{e}$ is the electronic momentum, $\hat{\bf J}^{v}$ is the vibrational momentum,
$R$ is the distance between Yb and the center mass of OH, $r$ is OH bond length
and $\theta$ is the angle between OH  and the axis ($z$ axis of the molecular frame) directed from Yb to the OH center of mass. The condition $\theta=0$ corresponds to the linear configuration where the O atom is between Yb and H ones. $R$, $r$ and $\theta$ are the so called Jacobi coordinates. $V(\theta)$ is the potential energy curve obtained in the electronic structure calculations \cite{Zakharova:21b}.

$ {\rm \bf\hat{H}}_{\rm hfs} $  and  ${\rm \bf\hat{H}}_{\rm S}$ are the hyperfine interaction with H nucleus and Stark interaction with the external electric field respectively. In the current work we have fixed parameters of the ${\rm \bf\hat{H}}_{\rm mol}$, ${\rm \bf\hat{H}}_{\rm hfs}$ and ${\rm \bf\hat{H}}_{\rm S}$ from Ref.~\cite{Petrov:2022} except electronic matrix element
\begin{multline}
\frac{1}{\mu R^2}
   \langle\Psi_{\Omega=1/2} |J^e_+|\Psi_{\Omega=-1/2} \rangle 
= p_0 + p_1P_{l=1m=0}(\theta),
\label{pme}
\end{multline}
where 
$\Psi_{\Omega}$ is the electronic wavefunction, 
$P_{lm}(\theta)$ is the associated Legendre polynomial.
In this paper we put $p_0 = 0.486  {\rm ~cm}^{-1}$ and $p_1 = 272  {\rm ~MHz}$. Using these parameters does not affect results of Ref. \cite{Petrov:2024} as well as parameters from \cite{Petrov:2024} can be used in this work which aim is to study $v=1$ state. Current parameters, however, together with $v=1$ correctly describe $v=0$ vibrational mode.

\begin{equation}
 {\rm \bf\hat{H}}_{\rm Z} = \mu_{\rm B}(\hat{ {\bf L}}^e-{\rm g}_{S}\hat{ {\bf S}}^e)\cdot{\bf B} -{\rm g}_{\rm H}\mu_{N}{\bf \rm I}\cdot{\bf B} -
 {\rm g}_{\rm v}\mu_{N}\hat{\bf J}^{v}\cdot{\bf B}
 \label{HZe}
\end{equation}
describes the interaction of the molecule with external magnetic field ${\bf B}$, ${\hat{ \bf L}}^e$ and ${\hat{\bf S}}^e$ are the electronic orbital and electronic spin momentum operators, respectively, ${\rm g}_{S} = -2.0023$ is a free$-$electron $g$-factor, $\mathbf{I}$ and $g_{\rm H} = 2.7928456$ are the
angular-momentum operator and g factor of the hydrogen nuclei, ${\rm g}_{\rm v}$ is the g factor associated with the vibrational momentum,
$\mu_{B}$ and $\mu_{N}$ are Bohr and nuclear magnetons respectively.

In the molecular frame coordinate system the Zeeman interaction with electrons (first term in the right hand side of Eq. (\ref{HZe})) are determined by the electronic matrix elements

\begin{eqnarray}
 \label{Gpar}
   \frac{1}{\Omega} \langle \Psi_{\Omega=1/2} |\hat{L}^e_z - g_{S} \hat{S}^e_z |\Psi_{\Omega=1/2} \rangle &=& G_{\parallel},  
\end{eqnarray}
\begin{eqnarray}
 \label{Gperp}
 \langle \Psi_{\Omega=1/2} |\hat{L}^e_{+} - g_{S} \hat{S}^e_{+} |\Psi_{\Omega=-1/2} \rangle &=& G_{\perp},  
\end{eqnarray}
\begin{eqnarray}
 \label{Gperp2}
 \langle \Psi_{\Omega=-1/2} |\hat{L}^e_{+} - g_{S} \hat{S}^e_{+} |\Psi_{\Omega=1/2} \rangle &=& \tilde{G}_{\perp}P_{l=2m=2}(\theta).  
\end{eqnarray}
In this paper the parameters  $G_{\parallel}=G_{\perp} = 2.07$ are taken from the experimental values for the effectiv g factor ${\rm g}_S =-2.07(2)$ for $v=1$ bending mode \cite{Jadbabaie_2023}. $\tilde{G}_{\perp} = 8.6\cdot10^{-3}$  is determined from approximated relation $\tilde{G}_{\perp} \approx p_2/2B_{\rm rot}$, where $B_{\rm rot}=7329$ MHz is the rotational constant \cite{Jadbabaie_2023}, $p_2 =  125.9   {\rm ~MHz}$ determines electronic matrix element $2B_{\rm rot} \langle\Psi_{\Omega=-1/2} |J^e_+ |\Psi_{\Omega=+1/2} \rangle  = p_2P_{l=2m=2}(\theta)$ \cite{Petrov:2024}. The value of ${\rm g}_{\rm v}$ will be discussed below.

Wavefunctions, rovibrational energies and hyperfine structure were obtained by numerical diagonalization of the Hamiltonian (\ref{Hamtot})
over the basis set of the electronic-rotational-vibrational-nuclear spins wavefunctions
\begin{equation}
 \Psi_{\Omega }P_{lm}(\theta)\Theta^{J}_{M_J,\omega}(\alpha,\beta)U^{\rm H}_{M^{\rm H}_I}.
\label{basis}
\end{equation}
Here 
 $\Theta^{J}_{M_J,\omega}(\alpha,\beta)=\sqrt{(2J+1)/{4\pi}}D^{J}_{M_J,\omega}(\alpha,\beta,\gamma=0)$ is the rotational wavefunction, $\alpha,\beta$ correspond to azimuthal and polar angles of the molecular $z$ axis  (directed from Yb
to the center mass of OH),
 $U^{\rm H}_{M^{\rm H}_I}$ is the hydrogen  nuclear spin wavefunction, $M_J$ is the projection of the molecular (electronic-rotational-vibrational) angular momentum $\hat{\bf J}$ on the lab axis, 
 $\omega$ is the projection of the same momentum on $z$ axis of the molecular frame,
 $M^{\rm H}_I$  is the projections of the nuclear angular 
momenta of hydrogen  on the lab axis,
$l$ is the vibration angular momentum and $m$ is its projection on the molecular axis.

In this  calculation functions with $\omega - m = \Omega = \pm 1/2$, $l=0-30$  and $m=0,\pm 1, \pm 2$, $J=1/2,3/2,5/2$  were included to the basis set (\ref{basis}).
The ground vibrational state $v=0$ corresponds to $m=0$,
the first excited bending mode $v=1$ (focus of this paper) to $m=\pm 1$, the second excited bending mode $v=2$ has states with $m=0, \pm2$ etc. 

%

\section{Results}
\subsection{Energy levels for field-free case}
For $e$EDM search the ground rotational (N=1) levels of the first excited $v=1$ bending vibrational mode of YbOH are considered. The levels are well described by the Hund's case $b$ coupling scheme \cite{Jadbabaie_2023}. Electron spin ${ S=1/2}$ for a good approximation is an integral of motion. Its interaction (spin-rotation) with the rovibrational momentum ${\bf N} = {\bf J} - {\bf S}$ gives rise to the splitting between the energy levels with total $J = 1/2$ and $J = 3/2$ momenta. The Coriolis interaction with 
$m=0$ modes
results in split of energies between $1 / \sqrt{2}(|m=+1\rangle+|m=-1\rangle)$ and $1 / \sqrt{2}(|m=+1\rangle -|m=-1\rangle)$ states having opposite parities.
This effect is known as  $l$-doubling (or parity doubling) and the corresponding states are the components of $l$-doublets. 
 $l-$doubling is, in general, different for the $J = 1/2$ and $J = 3/2$ levels. The experimental values of 35 MHz and 18.5 MHz for $J = 1/2$ and $J = 3/2$ respectively were obtained in Ref. \cite{Jadbabaie_2023}. The corresponding energy levels were described in Hund's case $b$ \cite{Jadbabaie_2023} as well as in Hund's case $c$ \cite{Petrov:2024} coupling schemes.

The effect of $l$-doubling is partially similar to $\Omega$-doubling one in diatomic molecules, though projection of the electronic momentum $\Omega = \pm 1$ should be used instead of projection of vibrational momentum $m = \pm 1$. The corresponding parity eigenstates have similar structure $1 / \sqrt{2}(|\Omega=+1\rangle \pm |\Omega=-1\rangle)$.

Hydrogen nucleus has a nonzero nuclear spin $I = 1/2$, which gives rise to
the hyperfine energy splitting (for each component of l-doublet) between the levels with total
(electronic-vibrational-rotational-nuclear spin) angular momentum $F = J \pm 1/2$. Finally we have two $F=0$, four $F=1$ and two $F=2$ hyperfine sublevels for $N=1$ rotational level. Since levels with projection of the total momentum
$M_F$ = 0 are insensitive to $e$EDM we have six $M_F$ = 1 and two $M_F$ = 2 levels which can be considered for $e$EDM measurement.


 \subsection{g factors}

We define the effective g factors such that Zeeman shift is equal to
\begin{equation} 
   E_{\rm Zeeman} = {\rm g}\mu_B B M_F.
 \label{Zeem}
\end{equation}


\begin{figure}
\includegraphics[width=0.95\linewidth]{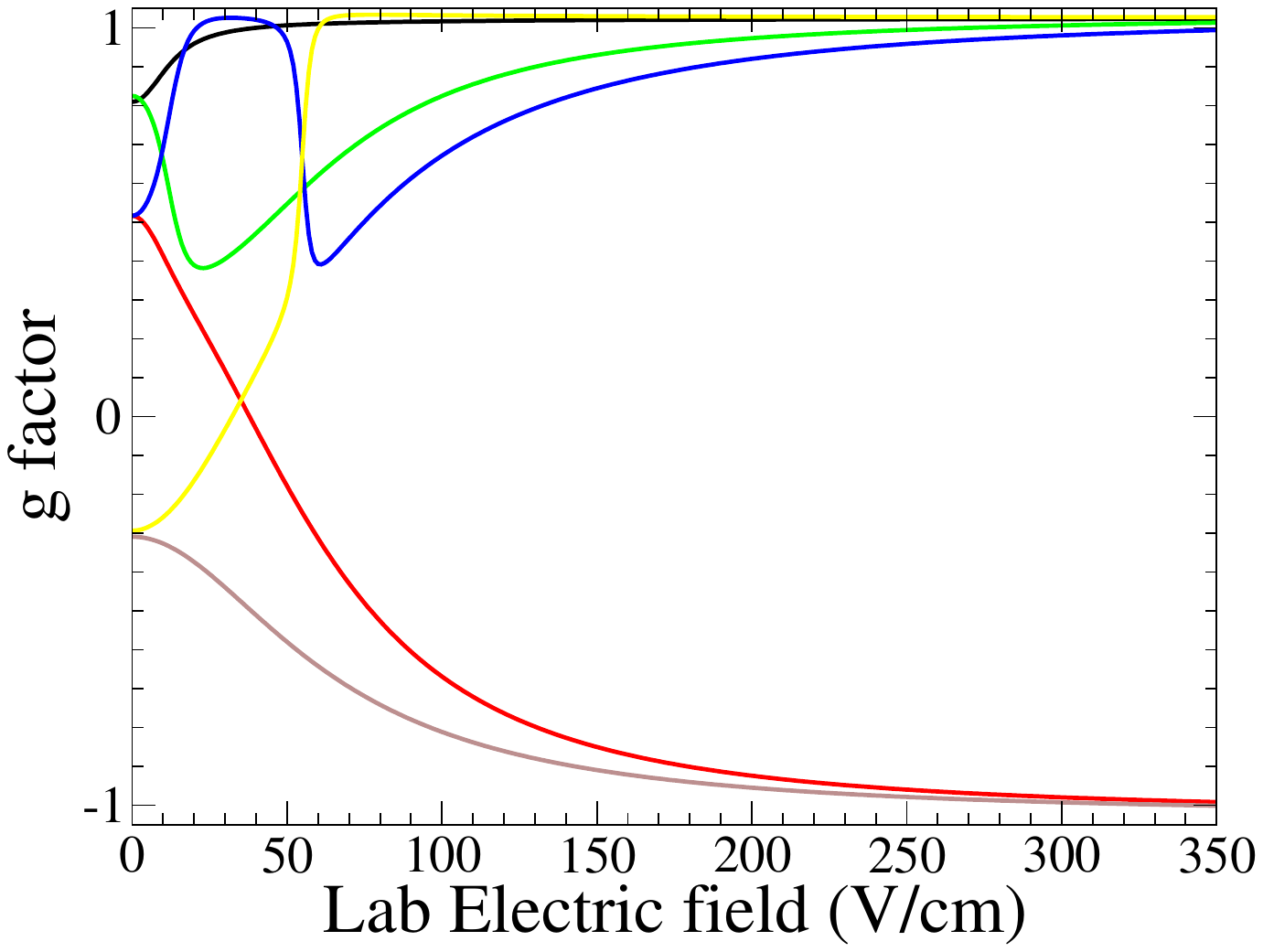}
 \caption{(Color online) Calculated g-factors 
 for the $N = 1, M_F=1$  Zeeman levels of $v=1$ vibrational state
 as functions of the laboratory electric field.}
 \label{gf1}
\end{figure}

\begin{figure}
\includegraphics[width=0.95\linewidth]{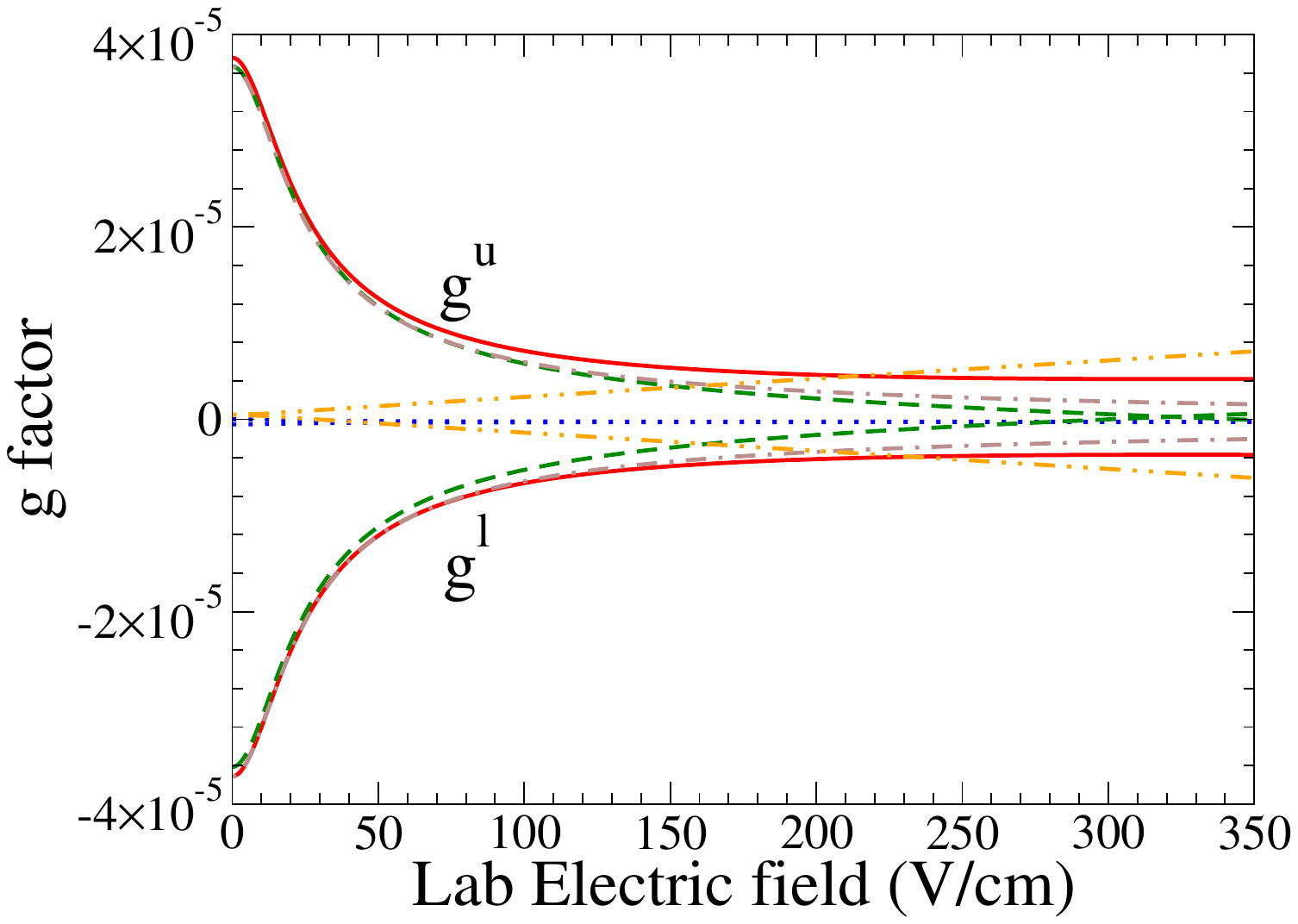}
 \caption{(Color online) Calculated g-factors for upper (${\rm g}^u$) and lower (${\rm g}^l$) components of $l$-doublet of $M_F=2$  Zeeman sublevels of the ground rotational $N = 1$ of the first excited $v=1$ vibrational mode as functions of the laboratory electric field.
 $G_{\parallel} = G_{\perp} = 2.07$ in calculations unless otherwise stated.
  Solid (red) lines: $\tilde{G}_\perp = 8.6\cdot 10^{-3}$, $ {\rm g}_v = 2$ in calculation, the shifted values ${\rm g} - 0.516471$ are plotted.
   Dashed (green) lines: $\tilde{G}_\perp = 8.6\cdot 10^{-3}$, $ {\rm g}_v = -2$ in calculation, the shifted values ${\rm g} - 0.517016$ are plotted.
    Dotted-dashed (brown) lines: $\tilde{G}_\perp = 8.6\cdot 10^{-3}$, $ {\rm g}_v = 0$ in calculation, the shifted values ${\rm g} - 0.516744$ are plotted.
 Dotted (blue) lines: $\tilde{G}_\perp = {\rm g}_v = 0$ in calculation, the shifted values ${\rm g}- 0.516744$ are plotted.
 Dotted-dotted-dashed (orange) lines:  $G_{\parallel} = 2.07$,   $G_{\perp} = 2.08$, $\tilde{G}_\perp = {\rm g}_v = 0$ in calculation, the shifted values ${\rm g}- 0.518245$ are plotted.} 
 \label{gf2}
\end{figure}

In Figure \ref{gf1} the calculated g factors for the $N = 1, M_F=1$ Zeeman levels of $v=1$ vibrational mode as functions of the laboratory electric field are given.
One can see that though there are points where g factors for some levels become equal, in general, g factors differences for $M_F=1$ levels due to complex interaction between them are rather large and irregular for electric field $E<$ 100 V/cm.

For $M_F=2$ Zeeman levels g factor difference is much smaller. Therefore we focus our attention to this case. We consider three main contributions to the difference. The first one is the interaction (\ref{Gperp2}). Due to the selection rules the operator
$\hat{L}^e_{+} - g_{S} \hat{S}^e_{+}$ has nozero matrix elements between $|\omega \rangle$ and $|\omega-1 \rangle$ states which for the case of matrix element (\ref{Gperp2}) implies interaction of the states with $m$ and $m-2$ projections of the vibrational momentum. These states just constitute components $1 / \sqrt{2}(|m=+1\rangle \pm |m=-1\rangle)$ of the $l$-doublet . Since $|m=+1\rangle $ and $|m=-1\rangle $ have opposite relative phases in different components of the $l$-doublet, accounting for interaction (\ref{Gperp2}) leads to opposite shifts for the g factors.
To the best of our knowledge matrix elements (\ref{Gperp2}) was never considered previously.

The second contribution is similar to that in diatomics.
Let us recall that
one should take into account the small admixture of $|\Omega=0 \rangle$ states to $\Omega$-doublet due to the Coriolis interaction. So the eigenfunctions of the $\Omega$-doublet components read $1 / \sqrt{2}(|\Omega=+1\rangle + |\Omega=-1\rangle) + \epsilon |\Omega=0^+\rangle$ for one component and $1 / \sqrt{2}(|\Omega=+1\rangle - |\Omega=-1\rangle) + {\epsilon}' |\Omega=0^-\rangle$ for another one, where $\epsilon \ll 1$ and $\epsilon' \ll 1$. Then interference between Coriolis and Zeeman interactions 
leads to a g factor difference. 
See e.g. Eqs. (12,13) in Ref. \cite{Petrov:14}. The same situation can be considered for $l$-doublets as well. However, since quantum number $m$ is associated with the vibrational momentum the Zeeman interaction with vibrational magnetic momentum should be considered, which is much smaller than the electronic one considered for $\Omega$-doublets.

To estimate the value of the vibrational g factor $ {\rm g}_v$  let us rewrite the magnetic moment of nuclear subsystem in space fixed frame
\begin{eqnarray}
\nonumber
    {\bf \mu }_n =   \frac{eZ_{\rm H}}{2c M_{\rm H}} \left[ {\bf r}_{\rm H}, -i\hbar\frac{ \partial}{\partial {\bf r}_{\rm H} }\right]_S  +  \\
    \frac{eZ_{\rm O}}{2c M_{\rm O}} \left[ {\bf r}_{\rm O}, -i\hbar\frac{ \partial}{\partial {\bf r}_{\rm O} }\right]_S +
     \frac{eZ_{\rm Yb}}{2c M_{\rm Yb}} \left[ {\bf r}_{\rm Yb}, -i\hbar\frac{ \partial}{\partial {\bf r}_{\rm Yb} }\right]_S
\label{munuc}
\end{eqnarray}
in Jacobi coordinates, keeping only terms with internal coordinates ${\bf r}$ and ${\bf R}$:
\begin{eqnarray}
\nonumber
    {\bf \mu }_n =  \frac{e}{2c} \left( \frac{Z_{\rm H}M_{\rm O}}{ M_{\rm H}\left(M_{\rm H} + M_{\rm O}  \right)} + 
     \frac{Z_{\rm O}M_{\rm H}}{ M_{\rm O}\left(M_{\rm H} + M_{\rm O}  \right)} \right) \times \\
\nonumber     
    \left[ {\bf r}, -i\hbar\frac{ \partial}{\partial {\bf r} }\right]_S  + \\
\nonumber    
  \frac{e}{2c} \left( \frac{(Z_{\rm H} + Z_{\rm O})M_{\rm Yb}}{\left(M_{\rm H} + M_{\rm O}  \right) M }  + 
 \frac{Z_{\rm Yb}\left(M_{\rm H} + M_{\rm O}  \right)}{M_{\rm Yb} M } 
        \right) \times \\  
\left[ {\bf R}, -i\hbar\frac{ \partial}{\partial {\bf R} }\right]_S,
\label{munucv}
\end{eqnarray}
where subscript S means that operators are taken in space fixed frame, $M = M_{\rm H} + M_{\rm O} + M_{\rm Yb}$.
Then recalling that  $\left[ {\bf r}, -i\hbar\frac{ \partial}{\partial {\bf r} }\right]_S = \left[ {\bf r}, -i\hbar\frac{ \partial}{\partial {\bf r} }\right]_B = \hat{\bf J}^{v}$, where subscript B means that operators are taken in the body (molecular) fixed frame and $\left[ {\bf R}, -i\hbar\frac{ \partial}{\partial {\bf R} }\right]_S = \hat{\bf J} -\hat{\bf J}^{e} - \hat{\bf J}^{v}$ obtain for the vibrational g factor
\begin{eqnarray}
\nonumber
    {\rm g }_v =   \left( \frac{Z_{\rm H}M_{\rm O}}{ M_{\rm H}\left(M_{\rm H} + M_{\rm O}  \right)} + 
     \frac{Z_{\rm O}M_{\rm H}}{ M_{\rm O}\left(M_{\rm H} + M_{\rm O}  \right)} \right) - \\
\nonumber    
   \left( \frac{(Z_{\rm H} + Z_{\rm O})M_{\rm Yb}}{\left(M_{\rm H} + M_{\rm O}  \right) M }  + 
 \frac{Z_{\rm Yb}\left(M_{\rm H} + M_{\rm O}  \right)}{M_{\rm Yb} M } 
        \right) = 0.45.
\label{gv}
\end{eqnarray}

At this value the vibrational magnetic moment gives only a small correction to the first contribution. Accounting for nonadiabatic interactions can potentially modify the ${\rm g}_v$ value, keeping it at the same order of magnitude however. To take this possible effect into account we performed calculations at slightly exaggerated values ${\rm g }_v = \pm 2$. Also we performed calculations when ${\rm g }_v =  0$ and when both $\tilde{G}_\perp =0$ and  ${\rm g }_v =  0$. Corresponding calculations are showed in Fig. \ref{gf2}. 

The third case leading to the g factor difference is the case when $G_{\parallel} \neq G_{\perp}$. It contributes only at nonzero electric field and will be discussed below together with the electric field dependence of the first two contributions.

Figure \ref{gf2} shows that g factors are almost coincide when  both $\tilde{G}_\perp =0$ and ${\rm g }_v = 0$ and
that the initial (at zero electric field) g factor difference for YbOH is mainly due to the first contribution (interaction (\ref{Gperp2})).  The external electric field mixes $l$-doublet components. Therefore,  one can expect that, when increasing the electric field, the initial small difference between g factors will gradually converge to zero. This is indeed the case when ${\rm g}_v = 0$ (Dotted-dashed (brown) lines in Fig. \ref{gf2}). This is not the case, however, when
${\rm g}_v \neq 0$. In this case the curves are similar for ones in diatomics (see e.g. Fig 1 in Ref. \cite{Petrov:11} or Fig. 2 in Ref. \cite{Petrov:14}). In full analogy to diatomics (as we already stated above, the way how vibrational momentum affects the g factor difference in linear triatomic molecules is similar to that how electronic one does in diatomics) to obtain the correct electric field dependence the perturbation from the excited $N = 2$ rotational level should be taken into account \cite{Petrov:11, Petrov:14}. Depending on the sign of the perturbation (sign of the ${\rm g}_v$ in our case) ${\rm g}^l$ and ${\rm g}^u$ do not tend to coincide (solid (red) line in Fig. \ref{gf2}, ${\rm g}_v = 2 > 0$) or the corresponding curves for ${\rm g}^l$ and ${\rm g}^u$ cross each other at electric field $E \approx 325$ V/cm and then diverge as well (dashed (green) line in Fig. \ref{gf2}, ${\rm g}_v = -2 < 0$).

If $G_{\parallel} = G_{\perp}$ then admixture by the external electric field of the exited $N=2$ rotational level does not affect the g factor difference. It can be seen from a simple consideration. The case $G_{\parallel} = G_{\perp}$ corresponds to Zeeman interaction of a free electron with an effectiv g factor $g_{s} = G_{\parallel}$ which, of course, is not affected by the external electric field. The g factors as functions of electric field for the case $G_{\parallel} = 2.07$ and $G_{\perp } = 2.08 $ (which corresponds to the error bar from Ref. \cite{Jadbabaie_2023}) in Fig. \ref{gf2} are plotted. One can see that the obtained g factor difference, by the order of magnitude, corresponds to the difference from other considered contributions.

From Fig. \ref{gf2} we can state finally that g factor difference of $l$-doublet components for $M_F=2$ Zeeman sublevels of $^{174}$YbOH is of order $\le 2\cdot 10^{-5}$ for the electric field in the range from 100 to 300 V/cm (comfortable for the experiment) and the ratio $\Delta {\rm g}/{\rm g} \le 4 \cdot 10^{-5}$ is more than order of magnitude smaller than the corresponding ratios $10^{-3}$ for the ThO \cite{Petrov:14} and HfF$^+$ \cite{ Petrov:17b}.

\section{Conclusion}
We determined main contributions leading to the g factor difference of $l$-doublet components in the first excited vibrational mode of linear triatomic molecules.
We found that g factors difference of $l$-doublet components for the $v=1, N = 1, M_F=2$ Zeeman sublevels of $^{174}$YbOH is of order $\simeq 2\cdot 10^{-5}$ for electric field in the range from 100 to 300 V/cm. The value is of key importance for the experiment for $e$EDM search on $^{174}$YbOH molecule.

\section{Acknowledgements}
Calculations of the g factor difference for $^{174}$YbOH are supported by the Russian Science Foundation grant no. 24-12-00092.
%
%

\end{document}